# NOMA Visible Light Communication System with Angle Diversity Receivers

**Mansourah K. Aljohani[1], Osama Zwaid Alsulami[1], Khulood D. Alazwary[1], Mohamed O. I. Musa[1], T.E.H.El-Gorashi[1], Mohammed T. Alresheedi[2] and Jaafar M.H. Elmirghani[1]**

[1] School of Electronic and Electrical Engineering, University of Leeds, Leeds, LS2 9JT, UK

[2] Department of Electrical Engineering, King Saud University, Riyadh, Kingdom of Saudi Arabia

E-mail: {ml16mka@Leeds.ac.uk, Ml15ozma@Leeds.ac.uk, elkal@Leeds.ac.uk, m.musa@leeds.ac.uk, t.e.h.elgorashi@leeds.ac.u,malresheedi@ksu.edu.sa, j.m.h.elmirghani@leeds.ac.uk

**ABSTRACT**

In this paper, a non-orthogonal multiple access (NOMA) visible light communication (VLC) system is investigated. The system uses angle diversity receivers (ADRs) to provide high data rates. The ADR has 4 branches, each directed to a different direction. An 8m × 4m sized room is modelled to study the resource allocation to users according to their channel conditions to maximize the data rate. The results show that using ADRs improves the data rate by an average of 35% compared to a system using wide FOV receivers.

## 1. INTRODUCTION

Demand for higher data rates has always been one of the main challenges in telecommunications. Researchers have been working on developing new approaches to provide high data rates for different applications. In Visible Light Communication (VLC), obtaining high data rates is one of the main challenges due to the limitations caused by the conventional transmitter and receiver technologies [1]-[11]. Other challenges include Inter Symbol Interference (ISI) and Co-Channel Interference (CCI) caused by the existence of multiple transmitters and multipath propagation. Laser diodes (LDs) are proposed to replace Light Emitting Diodes (LEDs) as transmitters because of their high modulation bandwidth which provides high data rates of up to several gigabits per second [12]-[14].

Different techniques were considered to enhance optical wireless communication systems including beam adaptation [15]–[17], spot diffusing approaches [18] – [23] and diversity techniques including angle diversity receiver (ADR) [24] - [27]. The use of code division multiple access (CDMA) schemes including pulse position modulation (PPM) CDMA, and multi-carrier CDMA (MC-CDMA) was investigated for multi-user communication [28] - [30]. Resource allocation in multi-user systems was investigated in [31] – [36].

To guarantee higher data rates, receivers such as Angle diversity receiver (ADR) are used [13], [17], [21] – [24]. An ADR is a combination of narrow Field of View (FOV) detectors that point to different directions to optimize the selection of the best and strongest optical signal. The use of ADRs aided in mitigating the effects of ambient light and pulse spread in optical wireless systems, as well as reducing the impact of interferences which improves the overall efficiency of the system [21]-[24]. Furthermore, ADRs can be used with other methods to improve the performance such as their use with relay nodes [37], fast adaptation techniques [38] – [40] and with beam steering in uplinks [41].

Non-Orthogonal Multiple Access (NOMA) in optical wireless systems operates by multiplexing all users' signals in the power domain then demultiplex them at the receiver end. It uses superposition coding for transmission and successive interference cancellation (SIC) at the receiving end. NOMA operates by allocating signal power according to the channel strength of each user; the higher their strength the lower the power assigned [42].

In this paper a NOMA system using ADRs with four branches is compared to a system using wide FOV receiver. Simulation models are developed to optimize the allocation of access point resources to users following our approach in [36]. The rest of the paper is organized as follows: The system model is introduced in Section 2. The simulation results and discussion are discussed in Section 3, while the conclusions are given in Section 4.

## 2. SYSTEM MODEL

The simulation models an empty room 8m long, 4m wide, and 3m heigh. Lambertian reflectors are used to model the walls and ceilings with a reflectivity coefficient of 0.8, without taking into consideration doors and windows [8], [9]. The signal is transmitted through multiple reflections from the walls and ceilings. Ray tracing is used for modelling the optical indoor channel similar to [11]-[13]. The reflective area is divided into several equal sized square shaped areas with area dA and reflection coefficient ρ. Reflections produced by surface elements reflecting the light emitted by a transmitter to the receiver are called first order reflections. The light reflected by a surface element can be further reflected by another surface element before reaching the receiver. This is called second order reflection. Reflections beyond the second order are insignificant due to their extremely low optical power [12]. The time resolution of the results is determined by the surface elements size. For first-order reflections, surface elements are of 5 cm × 5 cm while for second-order reflections, they are 20 cm × 20 cm. The dA values should be observed carefully due to their impact on the computation time.

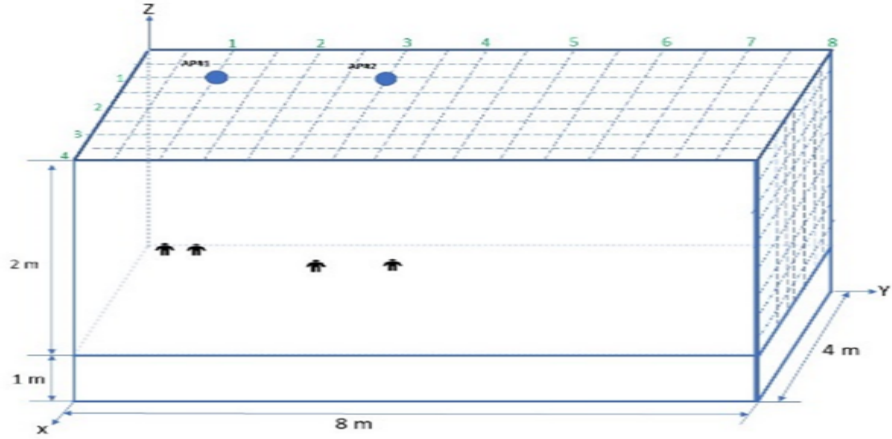

*Figure 1. Modelled Room dimensions*

To reduce interference, an ADR with 4 branches is used. Each branch has a photodetector that covers a different area based on the elevation and azimuth angles of the ADR faces, with the transmitter mounted in the ceiling. The elevation angles, $El$, of the four detectors are set at 70° whereas, the azimuth angles, $Az$, of the detectors are 45°, 135°, 225°, and 315°. The FOV of the four detectors is set to 25°. In addition, the area of each photodetector is 20 $mm^2$ with responsivity of 0.4 A/W [36].

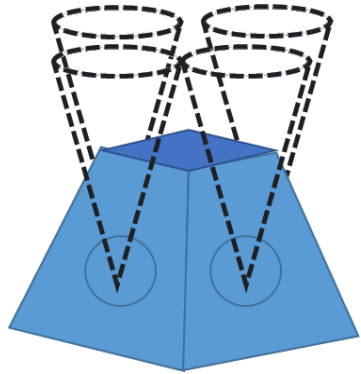

*Figure 2. Angle Diversity Receiver (ADR).*

We performed resource allocation to ensure that the sum SINR is maximized [36]. The power allocation coefficient for each user is similar to [43], [44]. The signal to interference ratio (SINR) for the NOMA system is given by:

$$SINR_k = \frac{(a_k P_t R h_k \eta)^2}{(\sum_{\substack{i=1 \\ i \neq k}}^{K} a_i P_t R h_k \eta)^2 + (BN_0 + 2q(I_d + R\, P_{bn})B)} \quad (1)$$

where:

| | |
|---|---|
| $a_k$ | Power allocation coefficient |
| $P_t$ | optical received power |
| R | Photodetector Responsivity |
| $h_k$ | Optical channel gain |
| $\eta$ | LD Efficiency |
| $N_0$ | Noise power density |
| B | Receiver bandwidth |
| q | Electron charge |
| $I_d$ | Dark background current |
| $P_{bn}$ | Received background optical power |

Table 1. shows a breakdown of all the system's configurations and characteristics, as well as the values used in the simulation and modelling of our design.

*Table 1, System Parameters*

| Parameters | Configurations | |
|---|---|---|
| Reflection coefficient of Walls and ceiling | 0.8 | |
| Reflection coefficient of Floor | 0.3 | |
| Reflections Count | 1 | 2 |
| Coverage Area | 5 cm x 5 cm | 20 cm x 20 cm |
| Reflection element angle | 60° | |
| Total transmit power | 1.9 W | |

## 3. SIMULATION RESULTS AND DISCUSSION

We compare the performance of VLC NOMA system using four branch ADR to the performance of a NOMA system using wide FOV receiver considering the room setup described in Section 2. The users in the room setup mentioned above are located in close proximity to the access points. The access points were located at (1,1,3), (1,3,3) while the users were located at (0.5,0.5,1), (0.5,1.5,1), (1.5,2.5,1), (1.5,3.5,1).

Figure 3, Figure 4 and Figure 5 show that using an ADR receiver instead of a wide FOV receiver is effective in increasing the data rate per user. For users located in close proximity to the access point, using the ADR (Figure 4) provides a range of improvements between 29% and 40% with an overall data rate improvement by an average of 35% compared to the data rate achieved by the wide FOV receiver (Figure 5). In general, the data rate in the wide FOV receiver system is lower due to the low channel bandwidth.

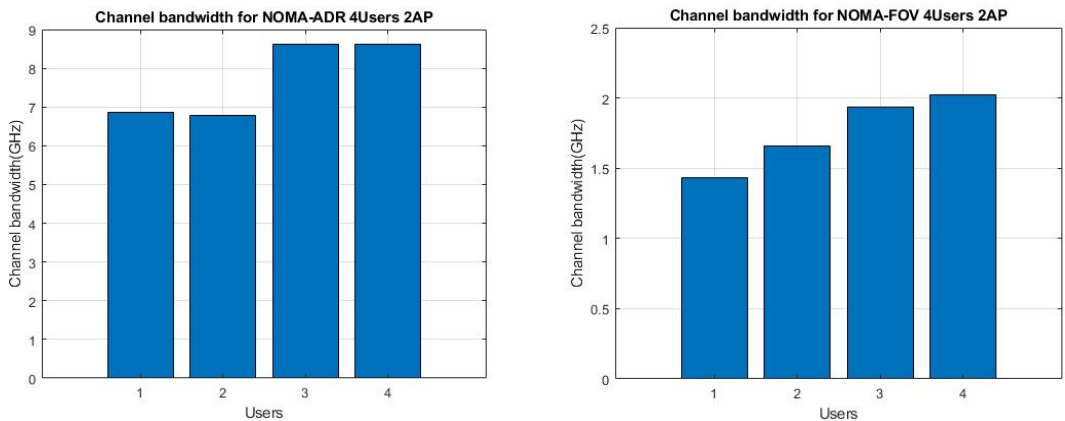

Figure 3. (a) Channel bandwidth (ADR). (b) Channel bandwidth (Wide FOV Receiver).

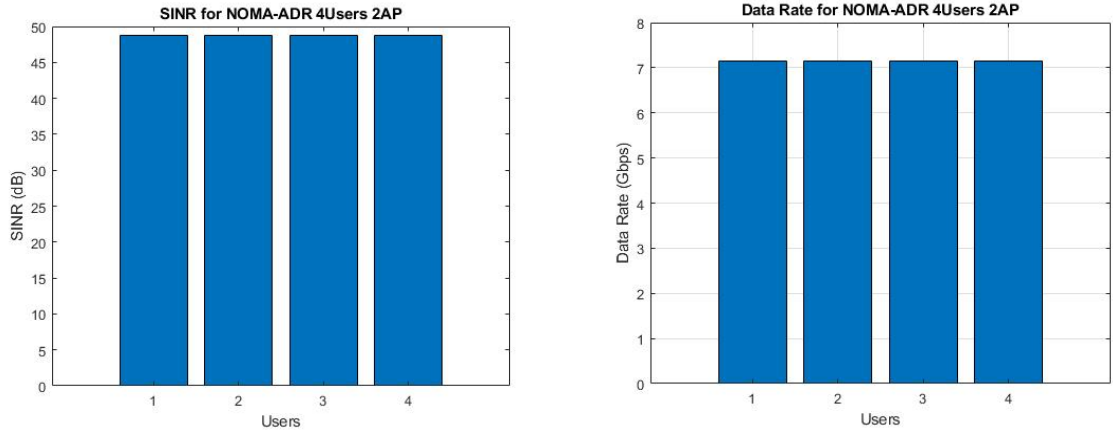
Figure 4. (a) SINR (ADR). (b) Data Rate (ADR).

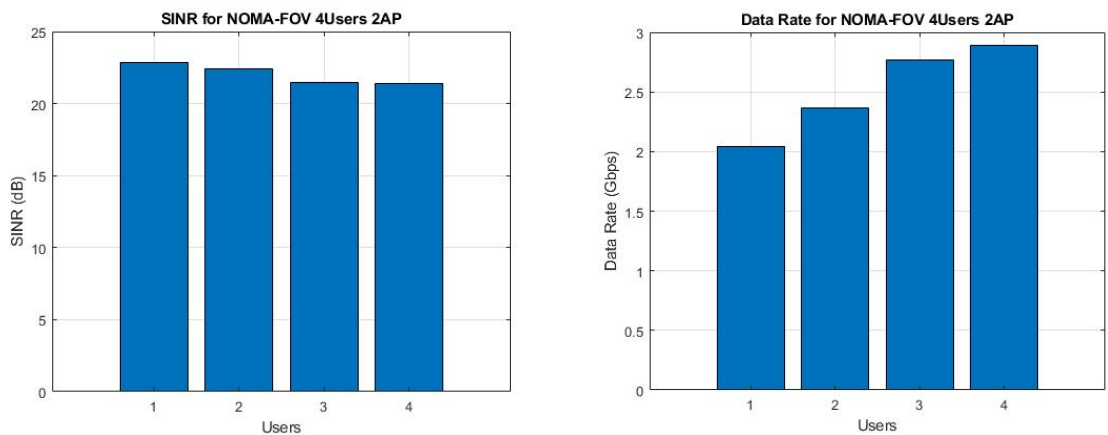
Figure 5. (a) SINR (Wide FOV Receiver). (b) Data Rate. (Wide FOV Receiver).

## 4. CONCLUSIONS

In this paper an ADR receiver with four branches was considered to enhance the data rate of a NOMA system. The performance of NOMA-VLC systems with an ADR is compared to a NOMA system with a wide FOV receiver in a room setup with two access points and multiple users in different locations. The simulation results revealed that using an ADR improves the data rate of the NOMA system by an average of 35% compared to a system with a wide FOV receiver.

**ACKNOWLEDGEMENTS**

The authors would like to acknowledge funding from the Engineering and Physical Sciences Research Council (EPSRC) INTERNET (EP/H040536/1), STAR (EP/K016873/1) and TOWS (EP/S016570/1) projects. The authors extend their appreciation to the deanship of Scientific Research under the International Scientific Partnership Program ISPP at King Saud University, Kingdom of Saudi Arabia for funding this research work through ISPP#0093. The first author would like to thank Taibah University, Kingdom of Saudi Arabia for funding her PhD scholarship. The second author would like to thank Umm Al-Qura University, Kingdom of Saudi Arabia for funding his PhD scholarship. All data is provided in the results section of this paper.

**REFERENCES:**
1. Fuada, Syifaul, et al. "A First Approach to Design Mobility Function and Noise Filter in VLC System Utilizing Low-cost Analog Circuits." *International Journal of Recent Contributions from Engineering, Science & IT (iJES) 5.2 (2017): 14-30.*
2. Cisco Visual Networking Index: Global Mobile Data Traffic Forecast Update, *2015–2020 White Paper,*
3. D. O'Brien, G. Parry, and P. Stavrinou, "Optical hotspots speed up wireless communication," *Nature Photonics, vol.1, no.5, pp.245-247, 2007.*
4. S. Arnon, "Visible light communication," *Cambridge University Press, 2015.*
5. Al-Ghamdi and J. M. Elmirghani, "Optimization of a triangular PFDR antenna in a fully diffuse OW system influenced by background noise and multipath propagation," *IEEE Transactions on Communications,* vol. 51, pp. 2103-2114, 2003

6. Al-Ghamdi and J. Elmirghani, "Performance analysis of mobile optical wireless systems employing a novel beam clustering method and diversity detection," *IEE Proceedings* in *Optoelectronics*, pp. 223-231, 2004
7. Alresheedi, Mohammed T., and Jaafar MH Elmirghani. "10 Gb/s indoor optical wireless systems employing beam delay, power, and angle adaptation methods with imaging detection." *Journal of Lightwave Technology* 30.12 (2012): 1843-1856
8. Barry, J.R., Kahn, J.M., Krause, W.J., Lee, E.A. and Messerschmitt, D.G., 1993. "Simulation of multipath impulse response for indoor wireless optical channels". IEEE journal on selected areas in communications, 11(3), pp.367-379.
9. F. E. Alsaadi and J. M. H. Elmirghani, "Mobile Multigigabit Indoor Optical Wireless Systems Employing Multibeam Power Adaptation and Imaging Diversity Receivers," *Journal of Optical Communications and Networking*, vol. 3, pp. 27-39, 2011.
10. H.H. Chan, H.H., K.L. Sterckx, J.M.H. Elmirghani, and R.A. Cryan, "Performance of optical wireless OOK and PPM systems under the constraints of ambient noise and multipath dispersion," *IEEE Communications Magazine*, Vol. 36, No. 12, pp. 83-87, 1998.
11. Alresheedi, Mohammed T., and Jaafar MH Elmirghani. "Hologram selection in realistic indoor optical wireless systems with angle diversity receivers." *Journal of Optical Communications and Networking* 7.8 (2015): 797-813.
12. Ahmed Taha Hussein, Mohammed T. Alresheedi, and Jaafar MH Elmirghani. "20 Gb/s mobile indoor visible light communication system employing beam steering and computer-generated holograms." *Journal of lightwave technology* 33.24 (2015): 5242-5260.
13. A. T. Hussein and J. M. Elmirghani, "10 Gbps mobile visible light communication system employing angle diversity, imaging receivers, and relay nodes," *Journal of Optical Communications and Networking*, vol. 7, pp. 718-735, 2015.
14. A. T. Hussein and J. M. H. Elmirghani, "Mobile Multi-Gigabit Visible Light Communication System in Realistic Indoor Environment," *Journal of Lightwave Technology*, vol. 33, pp. 3293-3307, 2015.
15. F. E. Alsaadi and J. M. H. Elmirghani, "Performance evaluation of 2.5 Gbit/s and 5 Gbit/s optical wireless systems employing a two dimensional adaptive beam clustering method and imaging diversity detection," *IEEE Journal on Selected Areas in Communications*, vol. 27, no. 8, pp. 1507–1519, 2009.
16. F. E. Alsaadi, M. Nikkar, and J. M. H. Elmirghani, "Adaptive mobile optical wireless systems employing a beam clustering method, diversity detection, and relay nodes," *IEEE Transactions on Communications*, vol. 58, no. 3, pp. 869–879, 2010.
17. M. T. Alresheedi and J. M. H. Elmirghani, "Performance evaluation of 5 Gbit/s and 10 Gbit/s mobile optical wireless systems employing beam angle and power adaptation with diversity receivers," *IEEE Journal on Selected Areas in Communications*, vol. 29, no. 6, pp. 1328–1340, 2011.
18. A. G. Al_Ghamdi and J. M. H. Elmirghani, "Line strip spot diffusing transmitter configuration for optical wireless systems influenced by background noise and multipath dispersion," *IEEE Transactions Communication, vol.52, no.1,pp.37 45, 2004.*
19. A. G. Al_Ghamdi and J. M. H. Elmirghani "Characterization of mobile spot diffusing optical wireless systems with diversity receiver," *IEEE International Conference in Communications (ICC), 2004.*
20. A. Al-Ghamdi, and J.M.H. Elmirghani, "Analysis of diffuse optical wireless channels employing spot diffusing techniques, diversity receivers, and combining schemes," *IEEE Transactions on communication*, Vol. 52, No. 10, pp. 1622-1631, 2004.
21. F. E. Alsaadi and J. M. H. Elmirghani, "High-speed spot diffusing mobile optical wireless system employing beam angle and power adaptation and imaging receivers," *IEEE/OSA Journal of Lightwave Technology*, vol. 28, no. 16, pp. 2191–2206, 2010.
22. A. G. Al-Ghamdi and J. M. H. Elmirghani, "Spot diffusing technique and angle diversity performance for high speed indoor diffuse infra-red wireless transmission," *IEE Proceedings Optoelectronics*, vol. 151, no. 1, pp. 46–52, 2004.
23. A. G. Al-Ghamdi and J. M. H. Elmirghani, "Characterization of mobile spot diffusing optical wireless systems with receiver diversity," *ICC'04 IEEE International Conference on Communications,* vol. 1, pp. 133-138, Paris, 20-24 June 2004.
24. A. T. Hussein and J. M. Elmirghani, "10 Gbps mobile visible light communication system employing angle diversity, imaging receivers, and relay nodes," *Journal of Optical Communications and Networking*, vol. 7, pp. 718-735, 2015.
25. A. Al-Ghamdi, and J.M.H. Elmirghani, "Performance evaluation of a triangular pyramidal fly-eye diversity detector for optical wireless communications," *IEEE Communications Magazine*, vol. 41, No. 3, pp. 80-86, 2003.
26. K.L. Sterckx, J.M.H. Elmirghani, and R.A. Cryan, "Sensitivity assessment of a three-segment pyrimadal fly-eye detector in a semi-disperse optical wireless communication link," *IEE Proceedings Optoelectronics*, vol. 147, No. 4, pp. 286-294, 2000.
27. K. L. Sterckx, J. M. H. Elmirghani, and R. A. Cryan, "Pyramidal fly-eye detection antenna for optical wireless systems," *Optical Wireless Communications. (Ref. No. 1999/128), IEE Colloq.*, pp. 5/1-5/6, 1999.
28. J. M. H. Elmirghani, and R.A. Cryan, "New PPM CDMA hybrid for indoor diffuse infrared channels," *Electron. Lett*, vol 30, No 20, pp. 1646-1647, 29 Sept. 1994.
29. F. E. Alsaadi and J. M. H. Elmirghani, "Adaptive mobile line strip multibeam MC-CDMA optical wireless system employing imaging detection in a real indoor environment," *IEEE Journal on Selected Areas in Communications*, vol. 27, no. 9, pp. 1663–1675, 2009.
30. F. E. Alsaadi and J. M. H. Elmirghani, "Adaptive mobile spot diffusing angle diversity MC-CDMA optical wireless system in a real indoor environment," *IEEE Transactions on Wireless Communications*, vol. 8, no. 4, pp. 2187–2192, 2009.
31. O. Z. Alsulami, M. T. Alresheedi, and J. M. H. Elmirghani, "Optical Wireless Cabin Communication System," in 2019 IEEE Conference on Standards for Communications and Networking (CSCN), 2019, pp. 1–4.
32. O. Z. Alsulami, M. O. I. Musa, M. T. Alresheedi, and J. M. H. Elmirghani, "Co-existence of Micro, Pico and Atto Cells in Optical Wireless Communication," in 2019 IEEE Conference on Standards for Communications and Networking (CSCN), 2019, pp. 1–5.
33. O. Z. Alsulami, M. O. I. Musa, M. T. Alresheedi, and J. M. H. Elmirghani, "Visible light optical data centre links," in 2019 21st International Conference on Transparent Optical Networks (ICTON), 2019, pp. 1–5.
34. S. O. M. Saeed, S. Hamid Mohamed, O. Z. Alsulami, M. T. Alresheedi, and J. M. H. Elmirghani, "Optimized resource allocation in multi-user WDM VLC systems," in 2019 21st International Conference on Transparent Optical Networks (ICTON), 2019, pp. 1–5.
35. Alsulami, O.Z., Aljohani, M.K., Saeed, S.O., Mohamed, S.H., El-Gorashi, T.E.H., Alresheedi, M.T. and Elmirghani, J.M., 2020. " Impact of room size on WDM optical wireless links with multiple access points and angle diversity receivers". *arXiv preprint arXiv:2002.09234.*
36. O. Z. Alsulami et al., "Optimum resource allocation in optical wireless systems with energy efficient fog and cloud architectures," Philosophical Transactions of the Royal Society A, Mathematics, Physics, Engineering and Science, vol. 378, No. 2169, pp. 1-11, March 2020.

37. Ahmed Taha Hussein and Jaafar M. H. Elmirghani. "10 Gbps mobile visible light communication system employing angle diversity, imaging receivers, and relay nodes." *Journal of Optical Communications and Networking* 7.8 (2015): 718-735.
38. Hussein, Ahmed Taha, Mohammed T. Alresheedi, and Jaafar M. H. Elmirghani. "Fast and efficient adaptation techniques for visible light communication systems." *IEEE/OSA Journal of Optical Communications and Networking* 8.6 (2016): 382-397.
39. F. E. Alsaadi, M. A. Alhartomi, and J. M. H. Elmirghani, "Fast and efficient adaptation algorithms for multi-gigabit wireless infrared systems," *IEEE/OSA Journal of Lightwave Technology*, vol. 31, no. 23, pp. 3735–3751, 2013.
40. Mohammed Thamer, Alresheedi, and Jaafar MH Elmirghani. "High-speed indoor optical wireless links employing fast angle and power adaptive computer-generated holograms with imaging receivers." IEEE Transactions on Communications vol. 64, No. 4 (2016): 1699-1710.
41. M. T. Alresheedi, A. T. Hussein, and J. M. Elmirghani, "Uplink design in VLC systems with IR sources and beam steering," *IET Communications*, vol. 11, pp. 311-317, 2017.
42. Almohimmah, E.M., Alresheedi, M.T., Abas, A.F. and Elmirghani, J., 2018, September. "A simple user grouping and pairing scheme for non-orthogonal multiple access in VLC system. "In *20th International Conference on Transparent Optical Networks (ICTON 2018). IEEE.*
43. Aljohani, M.K., Musa, M.O., Alresheedi, M.T. and Elmirghani, J.M., 2019, July. WDM NOMA VLC Systems. *In 2019 21st International Conference on Transparent Optical Networks (ICTON) (pp. 1-5). IEEE.*
44. Kizilirmak, Refik Caglar, Corbett Ray Rowell, and Murat Uysal. "Non-orthogonal multiple access (NOMA) for indoor visible light communications." *In 2015 4th international workshop on optical wireless communications (IWOW), pp. 98-101. IEEE, 2015.*